\documentclass[conference]{IEEEtran}
\IEEEoverridecommandlockouts
\usepackage{cite}
\usepackage{amsmath,amssymb,amsfonts}
\usepackage{algorithm}
\usepackage{algpseudocode}
\usepackage{graphicx}
\usepackage{textcomp}
\usepackage{xcolor}
\usepackage{xspace}
\usepackage{soul}
\usepackage{multirow}
\usepackage{booktabs}
\usepackage{caption}

\def\BibTeX{{\rm B\kern-.05em{\sc i\kern-.025em b}\kern-.08em
    T\kern-.1667em\lower.7ex\hbox{E}\kern-.125emX}}
\begin{document}

\title{A Reliable, Self-Adaptive Face Identification Framework via Lyapunov Optimization}

\author{
\IEEEauthorblockN{Dohyeon Kim}
\IEEEauthorblockA{\textit{Naver Webtoon}\\
Seongnam, Republic of Korea \\
dohyeonkim0412@gmail.com}
\and
\IEEEauthorblockN{Joongheon Kim}
\IEEEauthorblockA{\textit{Korea University} \\
Seoul, Republic of Korea \\
joongheon@korea.ac.kr}
\and
\IEEEauthorblockN{Jae young Bang}
\IEEEauthorblockA{\textit{Quandary Peak Research}\\
Los Angeles, CA, USA \\
jae@outlook.com}
}

\maketitle

\begin{abstract}
Realtime face identification (FID) from a video feed is highly computation-intensive, and may exhaust computation resources if performed on a device with a limited amount of resources (e.g., a mobile device). In general, FID performs better when images are sampled at a higher rate, minimizing false negatives. However, performing it at an overwhelmingly high rate exposes the system to the risk of a queue overflow that hampers the system's reliability. This paper proposes a novel, queue-aware FID framework that adapts the sampling rate to maximize the FID performance while avoiding a queue overflow by implementing the Lyapunov optimization. A preliminary evaluation via a trace-based simulation confirms the effectiveness of the framework.
\end{abstract}

\vspace{-3mm}

\section{Introduction}

%Every single day, brand-new machine learning algorithms have been investigated in many research areas such as computer vision, networks, systems, and big-data. The algorithms present excellent performances in terms of their own objectives. However, they can be quite computationally heavy to be operated in performance-limited computer systems such as mobile devices. Therefore, in the viewpoints of systems engineers, it is crucial to think about system-level supports for computation-rich learning algorithms. 

%This paper designs computer vision platforms which \textit{reliably} conducts face identification where the \textit{reliability} is defined as the avoidance of system queue overflow due to large computation delays in data-intensive applications. 

%Our considering platform is equipped with camera which samples images in real-time and conducts face identification. If the sampling frame rate is high (which is good for identifying fast moving human faces), it introduces delays due to large number of frames. This occurs queue-overflow that is obviously system-unstable (i.e., tradeoff between stability and identification performance). Thus, a new system-support algorithm which aims at time-average face identification performance maximization subject to system reliability under the concept of Lyapunov optimization framework. 

Realtime face identification (FID) is an active area of research in the artificial intelligence domain that has widely been adopted by the industry. As it matures, today, we are witnessing it spreading into our daily lives and being implemented even in mobile applications that often have limitations in the amount of computation resources available.

The computation resource limitation puts the system reliability and the FID performance in a trade-off relationship. An FID system samples images from a video feed and inserts them into a queue. The system concurrently retrieves the images and executes an FID algorithm to detect faces that appear in those images. With a higher sampling rate, the system would less likely to miss faces that do appear in the feed but not in the sampled images. However, an overwhelmingly high sampling rate could also cause a queue overflow, resulting in unexpected system behaviors. On the other hand, while sampling at a lower rate would help prevent an overflow, the system would more likely to fail to identify faces.

In this paper, we present a queue-aware FID framework that automatically adapts the sampling rate to the computation resource availability via the Lyapunov optimization in order to achieve maximum FID performance (fewest false negatives) while avoiding a queue overflow. We do not develop new FID algorithms; we exploit the existing ones and provide a self-adaptive framework on which the algorithms can be implemented and executed in a reliable fashion.

\begin{figure}[!t]
    \centering
    \includegraphics[scale=0.71]{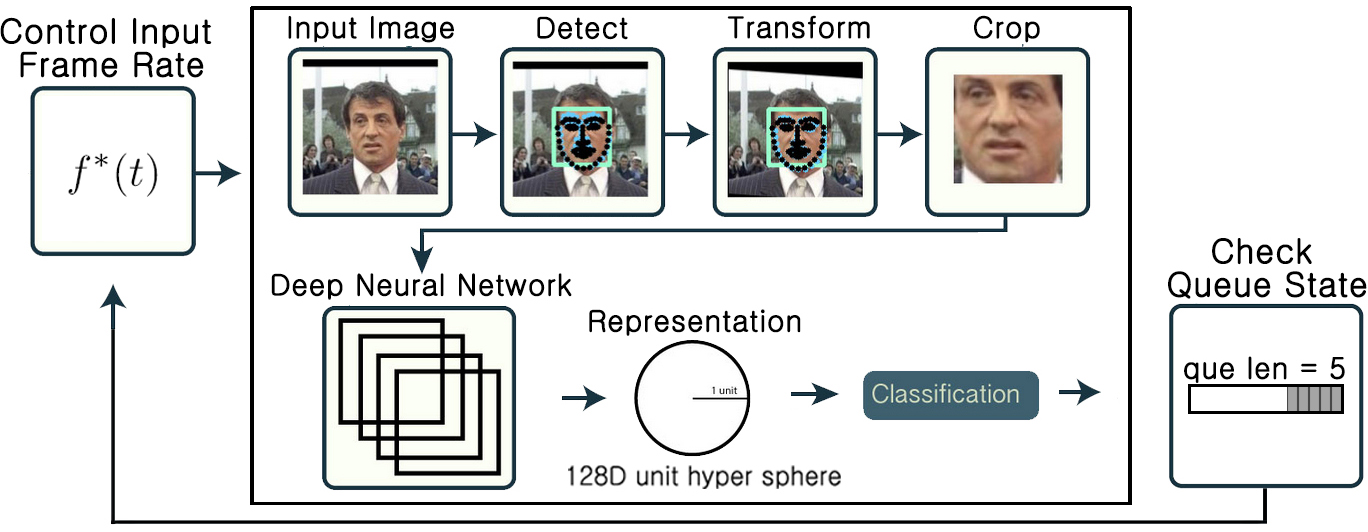}
    \caption{The proposed FID framework using OpenFace.}
    %Pipeline for the proposed face identification (FID) software platform with  OpenFace library.
    %for our basic reference system model. In here, we add additional steps on start and end of this system for frame rate adaptation
	\label{fig:pipeline_for_proposed}
	\vspace{-3mm}
\end{figure}

%\vspace{-3mm}
\section{Stochastic Frame Rate Adaptation}

%\vspace{0.3mm}
%\textbf{Lyapunov optimization framework:}

\subsection{Lyapunov optimization framework}

The theory of stochastic optimization~\cite{book2010sno} aims at optimizing a time-average utility subject to queue stability when the objective function and the queue stability constraints are in a trade-off relationship. Stochastic optimization models the queue stability using the Lyapunov drift. It takes actions that minimizes the Lyapunov drift while pursuing the minimization of a time-average objective function with the gap of $O(1/V)$ under queue stability with the bound of $O(V)$ %. By taking a control action, a time-average optimal utility can be obtained with the gap of $O(1/V)$ and a time-average queue-backlog bound of $O(V)$ 
where $V$ is defined as a trade-off between utility and stability. Neely discusses the details of the theory in his book~\cite{book2010sno}, and there are a number of applications that implement the theory~\cite{ton201608kim,tmc201907koo,jsac201806choi,jsac201811dao,twc201910choi,twc201912choi,twc202012choi,isj2021jung}.

%The theory of stochastic optimization aims at time-average optimization subject to queue/system stability when the objective function and queue stability constraints are in tradeoff relationship. 
%In the stochastic optimization formulation, Lyapunov control theory is used for the modeling of queue stability~\cite{book2010sno}. While pursuing the minimization of a time-average objective function, taking the minimum of the Lyapunov drift leads to the queue stability. 
%By taking a control action to minimize the both (time-average objective and drift) in each unit time, time-average optimal utility can be obtained with the gap of $O(1/V)$ from optimality while satisfying queue stability and a time-average queue backlog bound of $O(V)$ where $V$ is defined as a tradeoff factor between utility and queue stability. 
%More details about the theory of Lyapunov optimization are discussed in~\cite{book2010sno} and the applications of the theory are extensively studied~\cite{ton2016joongheon,tit2006neely,multimedia2017jonghoe}.

%\vspace{0.3mm}
%\textbf{Stochastic frame rate adaptation:}
%\vspace{-1mm}
\subsection{Reference face identification (FID) platform}
Fig.~\ref{fig:pipeline_for_proposed} depicts our reference FID framework. A system that implements the framework will have a video source (e.g., a camera) that constantly feeds a stream of images (frames) into the framework, and the framework stores the images in a queue in which they wait until the framework processes them. The framework concurrently conducts face identification on the images in the queue using the OpenFace library in the following four steps that correspond to those illustrated in Fig.~\ref{fig:pipeline_for_proposed}: (1) input image loading (\textit{Input Image}), (2) face recognition and preprocessing (\textit{Detect}, \textit{Transform}, and \textit{Crop}), (3) neural network forwarding (\textit{Deep Neural Network}, \textit{Representation}), and (4) classification.

The FID framework automatically controls the frame rate of the video feed in order to maximize FID performance while avoiding a queue overflow. In this paper, we define the \emph{FID performance} as following: the performance with a frame rate $f(t)$ at time $t$ is defined as $S(f(t))\triangleq\frac{\alpha(f(t))}{\beta(t)}$ where $\alpha(f(t))$ and $\beta(t)$ respectively denote the number of identified faces with given $f(t)$ and the total number of faces appeared in the original video feed, at time $t$. $S(f(t))=1$ if all faces are identified in $f(t)$ whereas $S(f(t))=0$ if no face is identified at all in $f(t)$.
%We make the assumption that the integrated FID algorithm has perfect recall (no false negative), which means a face is always identified if it appears in any of the frames. 
Higher $f(t)$ generally leads to higher $S(f(t))$; that lowers the chance of having faces that quickly pass by between two frames, unidentified (false negatives). However, in reality, because it is likely for the FID system to have a limited amount of queue storage and finite computation power, an overwhelmingly high $f(t)$ could cause a queue overflow and result in unreliable system behaviors. The framework continuously monitors the queue status and adapts the rate at which the frames from the video feed are inserted into the queue to the resource availability.

%Our considering reference FID software platform is as illustrated in Fig.~\ref{fig:pipeline_for_proposed}. Using the embedded camera in the FID platform, image frames are consistently arriving and the amount of arrived information depends on the frame rates, i.e., high frame rates introduce more information arrivals into the system. The low frame rates may lose fast moving human faces due to slow sampling rates, thus it leads to the degradation of system performance. In this paper, the system performance with frame rate $f(t)$ at time $t$, denoted by $S(f(t))$, defined as $S(f(t))\triangleq\frac{\alpha(f(t))}{\beta(t)}$ where $\alpha(f(t))$ and $\beta(t)$ stand for the number of identified faces with given $f(t)$ and the number of actual faces in the monitoring area, at time $t$. Therefore, $S(f(t))=1$ if all faces are identified with $f(t)$ whereas $S(f(t))=0$ if all faces are not identified at all with $f(t)$. It is also obvious that high $f(t)$ leads to high $S(f(t))$, i.e., $S(f(t))$ monotonously increases depending on $f(t)$. 

\begin{algorithm}[t]
%\small
\footnotesize
\caption{Frame rate control via Lyapunov optimization}
\label{alg:mindppr}
\begin{algorithmic}[1]
%\Statex $\hspace{-1.5em}\textbf{Initialize:}$
\State $Q(t)\leftarrow 0$ and $t\leftarrow 0$
%\State $Q_e \gets 0$, $Q_d \gets 0$, $Q_r \gets T_b/t_p$, and $N_{w,prev} \gets 0$
%\Statex $\hspace{-1.5em}\textbf{Main algorithm:}$
\While{$t\leq T$} // $T$: operation time
    \State Observe $Q(t)$ and $\mathcal{T}^{*} \leftarrow -\infty$
    \For{$f(t)\in\mathcal{F}$} 
            %\State 
            $\mathcal{T}
                \leftarrow V\cdot S(f(t))-Q(t)\cdot \lambda(f(t))$
            \If {$\mathcal{T} \geq \mathcal{T}^{*}$}
                %\State 
                $\mathcal{T}^{*}\leftarrow\mathcal{T}$ and $f^{*}(t)\leftarrow f(t)$
            \EndIf
        \EndFor
%        \State{Request video chunks according to $\pi^{opt}$}
\EndWhile
\end{algorithmic}
\end{algorithm}

\subsection{Frame rate control via Lyapunov optimization}

This section describes how we model the time-average optimal frame rate using the Lyapunov optimization~\cite{book2010sno} on which our FID framework controls $f(t)$ based. An FID system that implements our framework continuously adjusts $f(t)$ to the outputs of the optimal frame rate model.
%In our preference FID platform, image frames arrive into the queue in real-time via embedded camera modules. The arrival can be controlled by adapting frame rates $f(t)$ at $t$.
%The dynamics of the queue in Fig.~\ref{fig:pipeline_for_proposed} can be modeled as:

We first model the arrival queue dynamics as
%\begin{equation}
$Q(t+1)=\max\left\{Q(t)-\mu(t),0\right\}+\lambda(f(t))$
%\end{equation}
where
    $Q(t)$, $\mu(t)$, and $\lambda(f(t))$
    respectively denote
    the queue-backlog size,
    the number of images departing from the queue, and
    the number of images arriving in the queue with $f(t)$.%, i.e., the frame rate at time $t$.
We then formulate the mathematical program for maximizing the time-average system performance $S(f(t))$ as
%\begin{equation}
$\max: \lim_{t\rightarrow\infty}\sum_{\tau=0}^{t-1}\nolimits S(f(\tau))$
%\end{equation}
and the queue stability constraint (to avoid a queue overflow) as
%begin{equation}
$\lim_{t\rightarrow\infty}\frac{1}{t}\sum_{\tau=0}^{t-1}\nolimits Q(t)<\infty$.
%\end{equation}
According to the Lyapunov optimization theory~\cite{book2010sno}, this program can be re-formulated as following where $f^{*}(t)$ is time-average optimal frame rate for observed $Q(t)$:

\begin{equation}
f^{*}(t)\leftarrow \arg\max_{f(t)\in\mathcal{F}}\nolimits \left\{V\cdot S(f(t)) - Q(t)\cdot \lambda(f(t))\right\}
\label{eq:final}
\nonumber
\end{equation}

\section{Implementation and Evaluation}
%Our trace-based simulation showed that the proposed FID framework does achieve maximum FID performance without having a queue overflow. We built a prototype that implements the framework and deployed it on a personal laptop with 4 CPU cores, 8 GB of memory, and a GPU. We observed that the sampling rate can be up to 10 fps with built-in camera. Therefore, we control the rate from 1 to 10 based on queue-backlog. As presented in Fig.~\ref{fig:evaluation}, utilizing max static sampling rate (10 fps) for high performance leads to queue-divergence. On the other hand, static rate with high reliability (i.e., the lowest rate) guarantees queue stability with the minimum performance which is unfavorable. The proposed algorithms control the performance based on queue-backlog. If $V$ in (\ref{eq:final}) is high, it means performance is relatively important than stability, and vice versa. Therefore, the proposed algorithm with high $V$ pursues higher performance while sacrificing certain amounts of queue occupancy as observed.

\begin{figure}[t!]
    \centering
    \includegraphics[scale=0.32]{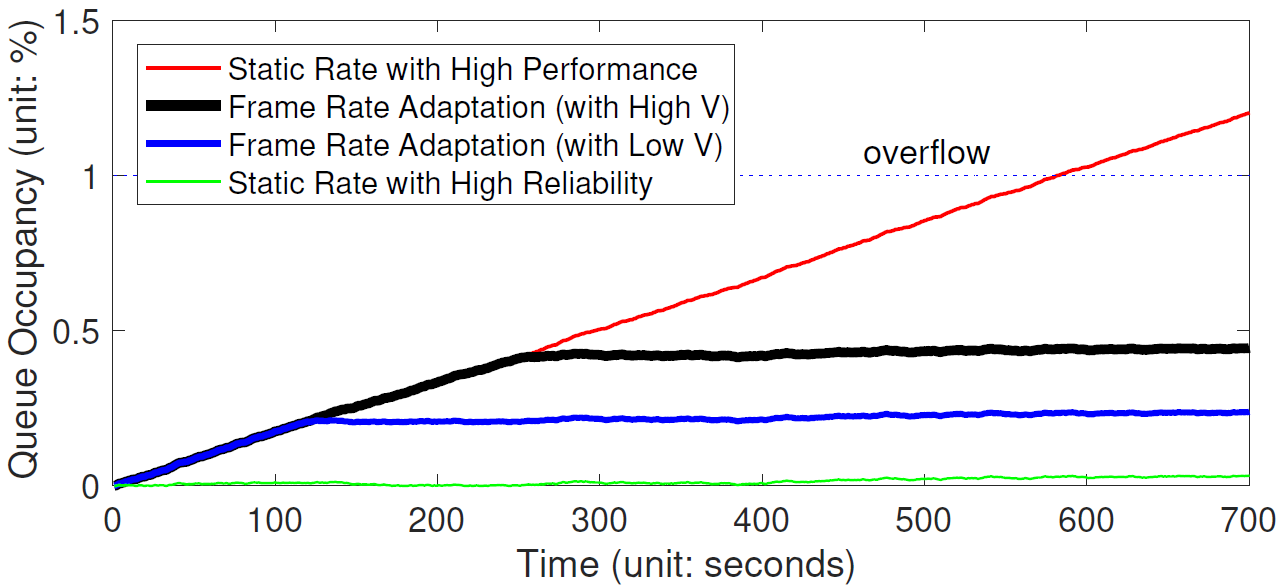}
    \caption{Evaluation results: queue dynamics.}
	\label{fig:evaluation}
	\vspace{-3mm}
\end{figure}

We evaluated our framework via a trace-based simulation, which showed that the framework maximizes the FID performance while not having any queue overflow. The evaluation was preliminary, and we made an assumption that maximizing the number of frames that the framework processes would also maximize the FID performance. We designed the simulation to mimic an FID system with a threshold of 10 frames/sec at which a queue-divergence would occur. We then varied the frame rate from 1 to 10, with and without our framework. Fig.~\ref{fig:evaluation} depicts the four simulation results: (1--red) the queue eventually overflows with a fixed frame rate of 10, (2--black / 3--blue) the queue stabilizes at certain points depending on the given $V$, and (4--green) the queue is stable but the FID performance is the lowest at 1 frames/sec. Our framework does not require a predetermined frame rate since it self-adapts the rate to the queue status on-the-fly.

%We built a prototype that implements the framework and deployed it on a personal laptop with 4 CPU cores, 8 GB of memory, and a GPU. We observed that the sampling rate can be up to 10 fps with built-in camera. Therefore, we control the rate from 1 to 10 based on queue-backlog. As presented in Fig.~\ref{fig:evaluation}, utilizing max static sampling rate (10 fps) for high performance leads to queue-divergence. On the other hand, static rate with high reliability (i.e., the lowest rate) guarantees queue stability with the minimum performance which is unfavorable. The proposed algorithms control the performance based on queue-backlog. If $V$ in (\ref{eq:final}) is high, it means performance is relatively important than stability, and vice versa. Therefore, the proposed algorithm with high $V$ pursues higher performance while sacrificing certain amounts of queue occupancy as observed.

\section{Concluding Remarks}
This paper proposes a reliable FID framework that achieves maximum FID performance by self-adapting the frame rate to the queue-backlog status using a unique Lyapunov optimization model. There are other system objectives such as time to identify a face or energy consumption of an FID system that would also be crucial for an FID system in practice, and we see an interesting research opportunity in creating new optimization models that target those objectives.

%As future research, the implemented software platform should be precisely evaluated with various system aspects.

\section*{Acknowledgment}
This research was supported by National Research Foundation of Korea (2017R1A4A1015675). J. Kim is a corresponding author.
This paper was presented at \textit{ACM Symposium on Operating Systems Principles (SOSP) Workshop on AI Systems (AISys)}, Shanghai, China, October 2017.


\begin{thebibliography}{3}
\bibitem{book2010sno} 
M. Neely, \textit{Stochastic network optimization with application to communication and queueing systems}. Morgan \& Claypool, 2010.

\bibitem{ton201608kim}
J. Kim, G. Caire, and A. F. Molisch, “Quality-aware streaming and scheduling for device-to-device video delivery,” \textit{IEEE/ACM Transactions on Networking}, vol. 24, no. 4, pp. 2319–-2331, August 2016.

\bibitem{tmc201907koo}
J. Koo, J. Yi, J. Kim, M. A. Hoque, and S. Choi, “Seamless dynamic adaptive streaming in LTE/Wi-Fi integrated network under smartphone resource constraints,” \textit{IEEE Trans. on Mobile Computing}, vol. 18, no. 7, pp. 1647-–1660, July 2019.

\bibitem{jsac201806choi}
M. Choi, J. Kim, and J. Moon, “Wireless video caching and dynamic streaming under differentiated quality requirements,” \textit{IEEE Journal on Selected Areas in Communications}, vol. 36, no. 6, pp. 1245-–1257, June 2018.

\bibitem{jsac201811dao}
N.-N. Dao, D.-N. Vu, W. Na, J. Kim, and S. Cho, “SGCO: Stabilized green crosshaul orchestration for dense IoT offloading services,” \textit{IEEE Journal on Selected Areas in Communications}, vol. 36, no. 11, pp. 2538-–2548, November 2018.

\bibitem{twc201910choi}
M. Choi, J. Kim, and J. Moon, “Dynamic power allocation and user scheduling for power-efficient and delay-constrained multiple access networks,” \textit{IEEE Trans. on Wireless Communications}, vol. 18, no. 10, pp. 4846--4858, October 2019.

\bibitem{twc201912choi}
M. Choi, A. No, M. Ji, and J. Kim, “Markov decision policies for dynamic video delivery in wireless caching networks,” \textit{IEEE Trans. on Wireless Communications}, vol. 18, no. 12, pp. 5705--5718, December 2019.

\bibitem{twc202012choi}
M. Choi, A. F. Molisch, and J. Kim, “Joint distributed link scheduling and power allocation for content delivery in wireless caching networks,” \textit{IEEE Trans. on Wireless Communications}, vol. 19, no. 12, pp. 7810--7824, December 2020.

\bibitem{isj2021jung}
S. Jung, J. Kim, and J.-H. Kim, “Intelligent active queue management for stabilized QoS guarantees in 5G mobile networks," \textit{IEEE Systems Journal}, vol. 15, no. 3, pp. 4293--4302, September 2021.
\end{thebibliography}
\end{document}